\documentclass[english]{llncs}

\usepackage[utf8]{inputenc}
\usepackage{babel}
\usepackage{verbatim}
\usepackage{csquotes}
\usepackage[final]{graphicx}
\usepackage{siunitx}
\usepackage{cite}
\usepackage[tworuled,linesnumbered,noend,noline]{algorithm2e}
\SetNlSty{textrm}{}{}

\usepackage{ifdraft}
\ifdraft{
  
}{
  
}

\usepackage{xspace}

\usepackage{amsmath}

\usepackage{amssymb}

\usepackage{booktabs}
\usepackage{multirow}
\usepackage{booktabs,caption,threeparttable}
\usepackage{tablefootnote}
\usepackage{tabularx}
\newcolumntype{W}{>{\raggedleft\arraybackslash}X}
\newcolumntype{T}{>{\centering\arraybackslash}X}

\usepackage[centerlast]{subfigure}

\newcommand{\titoletto}[1]{\smallskip\noindent\textbf{#1}~}

\usepackage[final]{hyperref}
\newcommand{\coll}[1]{{\small\url{#1}}}

\newcommand{\testotitolo}{%
  Using runs of homozygosity to detect genomic regions associated with
  susceptibility to infectious and metabolic diseases in dairy cows
  under intensive farming conditions
}
\newcommand{\testoautore}{%
  Filippo Biscarini
}
\newcommand{\testointest}{\testotitolo}

\hypersetup{%
  colorlinks=false,
  pdftitle    = {\testotitolo},
  pdfsubject = {},
  pdfkeywords = {},
  pdfauthor   = {\testoautore},
  pdfcreator = {Filippo Biscarini},
  pdfproducer = {},
  pdfborder = 0 0 0
}

\numberwithin{equation}{section}

\begin{document}

\title{\testotitolo}
\titlerunning{\testointest}
\author{%
  Filippo Biscarini\inst{1},
  Stefano Biffani\inst{2},
  Nicola Morandi\inst{1},
  Ezequiel L. Nicolazzi\inst{1} \and
  Alessandra Stella\inst{2}
}

\authorrunning{Biscarini \textit{et al.}}

\institute{%
Department of Bioinformatics, Parco Tecnologico Padano, Lodi, Italy,
\email{\{filippo.biscarini,nicola.morandi,ezequiel.nicolazzi\}@tecnoparco.org}
\and
IBBA-CNR, Lodi, Italy,
\email{\{biffani,stella\}.ibba.cnr.it}
}

\maketitle

\begin{abstract}
Runs of homozygosity (ROH) are contiguous stretches of homozygous genome
which likely reflect transmission from common ancestors and can be used
to track the inheritance of haplotypes of interest. In the present
paper, ROH were extracted from 50K SNPs and used to detect regions of
the genome associated with susceptibility to diseases in a population of
468 Holstein-Frisian cows. Diagnosed diseases were categorised as
infectious diseases, metabolic syndromes, mastitis, reproductive
diseases and locomotive disorders. ROH associated with infectious
diseases, mastitis and locomotive disorders were found on BTA 12. A long
region of homozygosity linked with metabolic syndromes, infectious and
reproductive diseases was detected on BTA 15, disclosing complex
relationships between immunity, metabolism and functional disorders. ROH
associated with infectious and reproductive diseases, mastitis and
metabolic syndromes were observed on chromosomes 3, 5, 7, 13 and 18.
Previous studies reported QTLs for milk production traits on all of
these regions, thus substantiating the known negative relationship
between selection for milk production and health in dairy cattle.
\end{abstract}



\section*{Introduction}


In dairy cattle intensive farming, diseases are certainly a major
concern. They reduce animal welfare, entail considerable direct
treatment costs, decrease milk production, may cause mortality and
present biosafety risks related to veterinary drugs (e.g. antibiotics)
residues in milk and the possible transmission to humans (zoonoses, e.g.
tuberculosis). 
Infectious (e.g. infectious mastitis) and metabolic (e.g. ketosis)
diseases comprise most of the diseases observed in dairy cattle. 

A strategy to reduce the incidence of diseases and improve the health
status of the herd could be to genetically enhance the resistance to
diseases. This could address both resistance to specific diseases and a
general improvement of the immune system of the animals ($\sim$ ``robustness
of dairy cows'' \cite{tenNapel2009}).

The identification of genomic regions associated with the susceptibility
to diseases is an important preliminary step towards implementing
breeding strategies aimed at improving dairy cattle health. This is now
made possible by the use of data obtained through next generation
genotyping and sequencing technologies (e.g. SNP chips, whole-genome
sequences). 
Genome-wide association studies (GWAS) are commonly used to scan the
genome in search of polymorphisms associated with the analyzed
phenotype. However, GWAS methods typically analyse one locus at a time,
are prone to spurious associations and, but for clear signals, are not
always of straightforward interpretation (e.g. \cite{McCarthy2008}).

Runs of homozygosity (ROH, \cite{McQuillan2008}) may be an alternative
approach to genome-wide scans for signals of genotype-phenotype
association. Runs of homozygosity are contiguous stretches of homozygous
genome which likely reflect transmission from a common ancestor, and can
therefore be considered as being IBD (identical by descent). ROH have
been used to estimate inbreeding both in humans \cite{McQuillan2008}
and cattle \cite{ferencakovic2011, purfield2012, kim2013}. Hildebrandt
et al. \cite{hildebrandt2009} applied 
a similar concept to map recessive disease genes in human populations,
and Biscarini et al. \cite{biscarini2013} used ROH to look for the
causal mutations for arthrogryposis and macroglossia in Piedmontese
cattle.

Hypothesizing that genetic variants associated with increased risk of
disease are more likely to be recessive than dominant (see Hildebrandt
et al. for human diseases \cite{hildebrandt2009}), looking for associations with
homozygous segments of the genome appears to be a reasonable strategy.
In this study, ROH were used to detect genomic regions associated with
susceptibility to 5 categories of diseases (infectious, metabolic and
reproductive diseases, mastitis and locomotive disorders) in dairy
cattle.

\section*{Material \& Methods}

\titoletto{Available data}

A population of 468 Holstein-Frisian cows between the first and fifth
lactation distributed over 4 herds (Table~\ref{tab:population} from the Po Valley region
in Northern Italy was analysed. All cows were farmed under intensive
conditions (use of concentrate feeds, no pasture, indoor housing) in
high-yielding dairy farms and were genotyped with the Illumina BovineSNP
50 beadchip version 2 (50k), based on the UMD 3.1 assembly of the \emph{Bos
taurus} genome.

\begin{table}[h!]
\centering
\caption{Population size after data editing. On the first row, per herd and total number of cows. Rows 2-6
  show n. of cases for each disease category$^1$. Disease category
  may overlap (e.g. infectious mastitis is both infectious disease and
  mastitis).}
      \begin{tabular}{cccccc}
        \hline
         & herd1& herd2 & herd3 & herd4 & total\\
        \hline
         COWS & 155 & 88 & 148 & 67 & 458\\
         INFD & 77 & 29 & 60 & 20 & 189\\
         METD & 62 & 10 & 68 & 10 & 152\\
         MAST & 46 & 13 & 44 & 13 & 117\\
         REPRD & 65 & 16 & 62 & 17 & 163\\
         LOCD & 21 & 6 & 33 & 13 & 74\\
         \hline
      \end{tabular}
      \begin{tablenotes}
      \small
      \item $^1$INFD: infectious
    diseases; METD: metabolic diseases; MAST: mastitis; REPRD:
    reproductive diseases; LOCD: locomotive disorders.
    \end{tablenotes}
     
      \label{tab:population}
\end{table}

Genotypic data were edited for individual and SNP
call rate ($>90\%$). Unmapped SNPs were removed, while those on the sex
chromosome were used. This left 458 cows and 53457 SNPs available for
the analysis.
Phenotype recording was carried out by veterinary practitioners within
the framework of the regional project Prozoo \cite{williams2011,
biscarini2012}: diagnosis, onset and treatment for each disease
were recorded. Given the limited number of cases for each specific
disease, diseases were grouped together in five homogeneous categories
in order to increase the statistical power of the analysis: infectious
diseases, metabolic syndromes, reproductive diseases, mastitis and
locomotive disorders. Categories were partially overlapping (for
instance, infectious mastitis was classified both as mastitis and
infectious disease). Table~\ref{tab:diseases} reports the classification of diagnosed
diseases into the above mentioned categories. There were 189, 152, 117,
163 and 74 cases respectively for infectious diseases, metabolic
syndromes, mastitis, reproductive diseases and locomotive disorders. For
each analysis, all animals not in the disease group were used as
controls (e.g. for infectious diseases there were 189 cases and 458 -
189 = 269 controls).

\begin{table}[h!]
\centering
\caption{List of diagnosed diseases falling in each of five
(partially overlapping) categories: infectious, metabolic
reproductive, locomotive diseases and mastitis.}
      \begin{tabular}{p{4cm} | p{6cm}}
        \hline
        \emph{Disease category} & \emph{Included diseases}\\
        \hline
          Infectious diseases & mastitis, peritonitis, enteritis,
          traumatic reticuloperitonitis, digital dermatitis,
          interdigital dermatitis, foot rot, laminitis, pyometra,
          metritis, endometritis, clostridiosis\\
        \hline
        Metabolic syndromes & milk fever, ovarian cysts, persistent
        corpus luteum, ruminal atony, displaced abomasum, indigestion,
        mesenteric torsion, volvulus, ketosis, steatosis\\
        \hline
        Reproductive diseases & retained placenta, dystocia, ovarian
        cysts, hypoplastic ovaries, hypotrophic ovaries, persistent
        corpus luteum, abortion, embryo resorption, metritis, mummified
        fetus, stillbirth, endometritis, parauterine abscess, pyometra\\
        \hline
        Mastitis & mastitis, teat obstruction, teat lesions\\
        \hline
        Locomotive disorders & digital dermatitis, interdigital
        dermatitis, sole ulcer, foot rot, white line disease, laminitis,
        tyloma, carpal arthritis, femoral fracture\\
        \hline
      \end{tabular}
      \label{tab:diseases}
\end{table}

\titoletto{Runs of homozygosity.}

Under the hypothesis that complex diseases have a genetic component made
up of several recessive variants distributed throughout the genome, each
with a small effect \cite{McQuillan2008}, runs of homozygosity (ROH)
were applied to detect genetic regions associated with susceptibility to
diseases. Single-SNP GWAS, which compares allele frequency at each
locus, would in fact detect also an excess of dominant alleles. ROH are
defined in diploid organisms as contiguous stretches of homozygous
genotypes, which reflect transmission of identical haplotypes from
common ancestors. Instead of focusing on a single locus, ROH consider
also the surrounding regions, thus accounting for the hitch-hiking
effect (neighbouring SNPs changing frequency together with a selected
locus with which they are in strong linkage disequilibrium;
\cite{barton2000}). The observed homozygosity was therefore estimated at each SNP
locus and averaged along a sliding window spanning 1000 kbps and
progressing SNP by SNP. A maximum of 5 missing genotypes and 1
heterozygous genotypes were permitted for a contiguous stretch of DNA to
be considered a ROH. 

ROH are basically a model-free statistical
technique. Unlike classic GWAS, there is no direct modeling of the
phenotype of interest nor a straightforward way to test for the strength
of detected associations. However, approaches can be conceived to assess
the significance of the detected signals. In this experimental
application of ROH to association studies, the significance of
phenotype/genotype associations was tested by looking at the difference
in homozygosity between cases and controls. The average homozygosity at
each SNP locus within the ROH was computed for cases and controls
separately, and the significance of the difference tested through a
one-tailed t-test: 

\[ 
\left\{
\begin{array}{l}
  H_0:\mu_{cases}=\mu_{controls}\\ 
  H_1:\mu_{cases}>\mu_{controls}
\end{array}
\right.
\]

Additionally, the resulting p-values were compared
with those obtained from a standard single-SNP GWAS for the same traits
on the same population. Drawing inspiration from the concept of
non-inferiority trials \cite{agostino2002}, the false discovery
rate (FDR) was computed for both sets of p-values and tested for
equivalence. The null hypothesis was that the FDR was actually larger
for ROH than in GWAS:

\[
H_0:FDR_{ROH} - FDR_{GWAS} > M
\]

where \emph{M} is a tolerance margin for the difference. This states that ROH are inferior to
GWAS. The alternative hypothesis was that the FDR was not different under both
methods:

\[
H_1:FDR_{ROH} - FDR_{GWAS} < M
\]

which implies that ROH are not inferior to GWAS. Non-inferiority was tested for the
5 traits analysed (infectious, metabolic, reproductive, locomotion
diseases and mastitis) with a tolerance margin $M = 0.01$. In all cases
$H_0$ was rejected, indicating that the ROH approach was not inferior to
standard GWAS (average p-value: $2.27 \cdot 10^{-6}$).

\titoletto{Software}

The software PLINK v1.07 \cite{purcell2007} was used for the analysis. No restriction on the minimum
number of SNPs in a ROH was applied, and the default (1000 kbps) maximum
gap between consecutive SNPs was used, in order to account for the lower
SNP density in the 50k SNP chip compared to the HD ($\sim777k$) SNP chip.
Data preparation, heritability estimates, graphical plots and
post-processing analyses were produced within the open source
programming environment R \cite{R2013}.

\section*{Results \& Discussion}

A total of 1273 distinct ROH were detected: this number comprised all
ROH found with the chosen parameters, irrespective of whether they were
found in all cows or a subset of them. The average ROH length was 285.7
kbps. In principle, stretches of homozygous DNA can appear in all
animals, regardless of their health status; those associated with
susceptibility to disease are however supposed to be more frequent in
cases than in controls. Therefore, the longest ROH (less likely to be
due to chance) which were most frequent in cases were retained as
results of interest and are reported in Table~\ref{tab:associations}. These include 17 ROH
for infectious diseases, 10 for reproductive diseases, 7 for metabolic
syndromes and 4 for mastitis and locomotive disorders. For the 42
reported ROH, the frequency in cases relative to controls varied from
$51.4\%$ to $100\%$; their average p-value and FDR were $0.398$ and $0.612$,
respectively. ROH in Table 3 were not always observed in all cases and
controls: the number of cows (cases + controls) involved ranged from 124
for infectious and metabolic disease to 56 for reproductive diseases. An
important issue in association studies is the expected statistical power
of the analysis. This is known to depend on sample size, the
heritability of the trait and linkage disequilibrium (LD) between
markers and QTLs \cite{luo1998}. Heritabilities for the diseases traits
included in this study were estimated with a sire threshold model for
binary traits fitting a random sire effect and different sets of
systematic effects ($herd, herd + calving\_month, herd + calving\_month +
age$). Heritability was not always estimable for all traits and models;
when estimable, it ranged between 0.05 and 0.10 (in line with literature
values: e.g. \cite{cole2013}). The average LD between adjacent markers
in the available Holstein-Frisian population was estimated as $r^2=0.23$.
With 458 individuals, the power to detect a QTL explaining $1\%$ of the
phenotypic variance would be therefore about $18\%$.
 
In the next section a few results of interest are described. The reader is warned
that, given the low power of the study and the high FDR for
all reported associations, this mostly has only speculative value. It is
however indicative of the output obtained from a ROH analysis.

\begin{table}[h!]
\centering
\caption{Most relevant runs of homozygosity (ROH) associated with the five disease categories considered in the
analysis. For each ROH, the chromosome (BTA), start and end position and number of SNPs included are reported. In
the reported ROH the percentage of cases on total number of haplotypes ranged between $51.4\%$ and $100\%$.}
      \begin{tabular}{lrrrrl}
        \hline
          \emph{Disease category} & \emph{BTA} & \emph{start bps} & \emph{end bps} & \emph{\# SNPs} &
          \emph{Production QTLs $^1$}\\
        \hline
         Infectious & 7 & 2991449 & 4655753 &47 & \\
         Infectious &7 &19163934 &19461567 &6 &FY, NM\\
         Infectious &7 &21704630 &22305419 &13 &FY, NM\\
         Infectious &7 &23756162 &24014128 &6 &FY, NM\\
         Infectious &7 &24233634 &24330857 &3 &FY, NM, PY\\
         Infectious &7 &106927241 &107452906 &12 &NM\\
         Infectious &7 &108575314 &108685133 &3 &NM\\
         Infectious &11 &45688827 &51041758 &117&\\ 
         Infectious &12 &11414743 &13005009 &43&\\
         Infectious &12 &25878820 &30099199 &84&\\ 
         Infectious &12 &51834816 &52573538 &21&\\
         Infectious &12 &53241975 &53428996 &6&\\
         Infectious &15 &40889434 &41169953 &4&\\
         Infectious &15 &41780731 &41971248 &7&\\
         Infectious &15 &42715966 &42758371 &2&\\
         Infectious &24 &16012254 &16274083 &8&\\
         Infectious &28 &30070377 &30158724 &2&\\
         Metabolic &3 &1737421 &6813316 &103&MY, FP\\ 
         Metabolic &4 &12868626 &13148484 &4&\\
         Metabolic &8 &8082781 &9666303 &37&\\
         Metabolic &15 &37893014 &39016114 &22&\\
         Metabolic &15 &40889434 &41169953 &4&\\
         Metabolic &15 &41780731 &41971248 &7&\\
         Metabolic &15 &42715966 &42758371 &2&\\ 
         Mastitis &12 &25878820 &30099199 &84&\\
         Mastitis &13 &19816277 &20687500& 22& MY, FY\\
         Mastitis &24 &16012254 &16274083 &8&\\
         Mastitis &28 &30070377 &30158724 &2&\\
         Reproduction &4 &8779979 &10737673 &40&\\
         Reproduction &4 &12868626 &13148484 &4&\\
         Reproduction &5 &6656617 &6976839 &11 &FY, PY\\
         Reproduction &5 &7832521 &8063248 &9 &FY, PY\\
         Reproduction &15 &40889434 &41169953 &4&\\
         Reproduction &15 &41780731 &41971248 &7&\\
         Reproduction &15 &42715966 &42758371 &2&\\
         Reproduction &18 &23224334 &23253048 &2&MY, FP\\ 
         Reproduction &20 &19587484 &19757303 &3&\\
         Reproduction &25 &30824927 &31525961 &27&\\
         Locomotion &6 &107186270 &107835731 &11&\\
         Locomotion &9 &78576880 &79887463 &14&\\
         Locomotion &12 &11414743 &13005009 &43&\\
         Locomotion &28 &30070377 &30158724& 2&\\
         \hline
      \end{tabular}
      \begin{tablenotes}
      \small
      \item $^1$QTLs for milk production traits from previous GWAS studies: FY=fat yield, NM=net merit, PY=protein yield,
MY=milk yield, FP=fat percentage (Cole et al., 2011; Minozzi et al.,
2013).
    \end{tablenotes}
      \label{tab:associations}
\end{table}

\titoletto{Regions associated with disease susceptibility}

\begin{figure}[!h]
\centering
\includegraphics[width=0.9\textwidth]{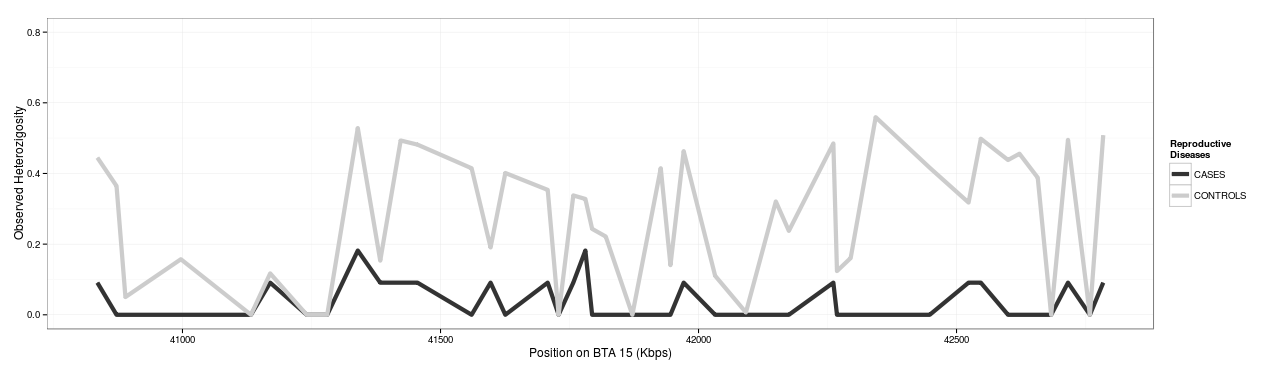}
\caption{Observed heterozygosity in cases (black line) and controls (grey line) for reproductive diseases along chromosome 15.}
\label{fig:roh}
\end{figure}

The longest ROH (117 SNPs, $\sim5.3$ Mbps) was found on BTA 11 and was
associated with infectious diseases. On BTA 12, three distinct regions
(\SI{11414743} - \SI{13050009} bps; \SI{25878820} - \SI{30099199} bps; \SI{51834816} -
\SI{53428996} bps) were associated with infectious diseases. The first two
of these regions were also found in cows with, respectively, locomotive
disorders and mastitis. Both conditions often have a microbial etiology
(e.g. infectious mastitis, foot rot). Interestingly, the ROH region at
$\sim12$ Mbps on BTA 12 contains the VWA8 gene (von Willebrand factor A
domain), whose mutations might be implied in coagulation abnormalities
\cite{sullivan1994} and may be related with musculoskeletal disorders.
Two other regions linked with both infectious diseases and mastitis were
found on BTA 24 and BTA 28. On BTA 28 Kolbehdari et al.
\cite{kolbehdari2008} found polymorphisms (\SI{28149059} bps and \SI{24423785}
bps) associated with the conformation of the mammary system and
angularity, which might be related to the occurrence of mastitis. A poor
conformation of the udder may be a predisposing factor for mastitis, as
well as intensive selection for dairy type. The ROH on BTA 28 was found
to be associated also with locomotive disorders, pointing at possible
detrimental effects of selection for dairy type on locomotion. 
Homozygous haplotypes linked with locomotive disorders were detected
also on BTA 6 between \SI{107186270} and \SI{107835761} bps; Kolbehdari et al.
\cite{kolbehdari2008} reported a nearby association with overall rump conformation,
which can be related to locomotion anomalies. Upstream along BTA 6 the
casein ($\alpha$-s1, $\alpha$-s2, $\beta$ and $\kappa$) loci are located which could
hint at a link between selection for casein variants and general
functionality of the animals.\\
In some cases, consecutive ROH are quite
close together (e.g. BTA 7 and BTA 15) which might reflect the existence
of a founder allele broken down by recombination over generations. The
parameters for ROH detection (consecutive SNP gap $>$ 1Mbps, more missing
or heterozygous SNPs) may also play a role, though. 
Figure~\ref{fig:roh} shows the
average observed heterozygosity in cases and controls for reproductive
diseases along BTA 15: ROH are visible as regions of low heterozygosity
(and conversely high homozygosity) in cases (black line) as compared to
controls (grey line). ROH associated with impaired reproductive function
were found on chromosomes 4, 8, and 18, where other studies detected
associations with calving ease \cite{kolbehdari2008}, calving
interval \cite{minozzi2013} and type traits like body depth, rump width,
stature and strength \cite{cole2009}.\\
In this study we identified
genetic associations with infectious diseases, mastitis, metabolic
syndromes and reproductive diseases on chromosomes 3, 5, 7, 13 and 18
where earlier works reported QTLs for milk production traits
\cite{cole2011, minozzi2013}: this confirms the known negative
relationship between strong selection for milk production and the health
of the animals. Negative genetic correlations between milk production
and disease resistance in Holstein (e.g. \cite{hansen2002}) and dairy
cattle in general (Norwegian Red, \cite{simianer1991} were estimated.
Though negative, such correlations are larger than -1; this implies that
alleles at polymorphic loci with positive effects on both
characteristics (or positive effects on production and neutral effect on
health) can be found (see for instance Pimentel et al.
\cite{pimentel2010} for fertility and production).

\section*{Conclusions}

Scanning the genome for runs of homozygosity provides a valid alternative to GWAS studies
for the genetic dissection of complex disease traits and for detecting
such variants with beneficial effects on both production and the health
of the animals. Results could be used in breeding programs aimed at
improving milk production while enhancing the resistance to diseases at
the same time. Emphasis in selection objectives could be more strongly
placed on functional rather than productive traits, thus improving
functionality while not completely neglecting production, thanks to
residual positive correlations. Alternatively, haplotypes favorably
linked to either production, functionality or both could be used to
identify carrier animals to breed next generation, in a sort of
``haplotype selection''. Several aspects of the use of ROH in association studies still need to
be investigated, though. Examples are: the construction of significance
tests (e.g. here the difference in homozygosity between cases and
controls was tested; however, the differential frequency of ROH could be
tested instead); the issue of multiple testing (e.g. permute cases and
controls so to create a distribution of ROH under the hypothesis of no
association); accounting for systematic effects (e.g. using residuals
instead of the original observations, or performing the ROH analysis
within class of effects); accounting for selection bias due to culling
of animals for health or productive reasons (e.g. ROH analysis within
parity/age class, or application of survival analysis). This promises to
be an exciting area of research: hopefully, this communication might
serve of inspiration.

\section*{Acknowledgments}

The results presented in this paper were produced within the framework
of the project "ProZoo" (www.tecnoparco.org/prozoo), financed by
"Regione Lombardia", "Fondazione Cariplo" and "Fondazione Banca Popolare
di Lodi". This research was also supported by the Marie Curie European
Reintegration Grant "NEUTRADAPT" and the "Gene2Farm" project within the
7th European Community Framework Programme, and by the Italian national
project "InterOmics" (www.interomics.eu).


\begin{thebibliography}{10}
\bibitem{tenNapel2009}
J. ten Napel, M.P.L. Calus, H.A. Mulder and R.F. Veerkamp.
\newblock Genomic concepts to improve robustness of dairy cows.
\newblock {\em EAAP publication}, 126:35--43, 2009.

\bibitem{McCarthy2008}
M.I. McCarthy, G.R. Abecasis, L.R. Cardon, D.B. Goldstein, J. Little,
J.P.A. Ioannidis and J.N. Hirschorn.
\newblock Genomic concepts to improve robustness of dairy cows.
\newblock {\em Nat. Rev. Genet.}, 9(5):356--369, 2008.

\bibitem{McQuillan2008}
R. McQuillan, A.L. Leutenegger, R. Abdel-Rahman, C.S. Franklin, M.
Pericic, L. Barac-Lauc, N. Smolej-Narancic, B. Janicijevic, O. Polasek,
A. Tenesa, A. MacLeod, S.M. Farrington, P. Rudan, C. Hayward, V. Vitart,
I. Rudan, S. Wild, M. Dunlop, A. Wright, H. Campbell and J. Wilson.
\newblock Runs of homozygosity in European populations.
\newblock {\em Am. J. Hum. Genet.}, 83:359--372, 2008.

\bibitem{ferencakovic2011}
M. Feren\v{c}akovi\'{c}, E. Hamzic, B. Gredler, I. Curik and J. S\"{o}lkner.
\newblock Runs of homozygosity reveal genome-wide autozygosity in the
Austrian Fleckvieh cattle.
\newblock {\em Agric. Consp. Sci.}, 76(4), 2011.

\bibitem{kim2013}
E.S. Kim, J.B. Cole, H. Huso, G.R. Wiggans, C.P. Van Tassell, B.A. Crooker, G. Liu, Y. Da and T.S. Sonstegard.
\newblock Effect of artifical selection on runs of homozygosity in U.S.
Holstein cattle.
\newblock {\em PLoS One}, 8(11):e80813, 2013.

\bibitem{purfield2012}
D.C. Purfield, D.P. Berry, S. McParland and D.G. Bradely.
\newblock Runs of homozygosity and population history in cattle.
\newblock {\em BMC Genetics}, 13:70, 2012.

\bibitem{hildebrandt2009}
F. Hildebrandt, S.F. Heeringa, F. Rüschendorf, M. Attanasio, G.
Nürnberg, C. Becker  ... E.A. Otto.
\newblock A systematic approach to mapping recessive genes in
individuals from outbred populations.
\newblock {\em PLoS Genetics}, 5(1):e1000353, 2009.

\bibitem{biscarini2013}
F. Biscarini, M. Del Corvo, A. Stella, A. Albera, M. Feren\v{c}akovi\'{c} and G. Pollott.
\newblock Busqueda de las mutaciones causales para artrogriposis y
macroglosia en vacuno de raza Piemontesa: resultados preliminares.
\newblock {\em Actas de las XV Jornadas sobre Producci\'{o}n Animal - AIDA}, 538-540, 2013.

\bibitem{williams2011}
J.L. Williams, S. Biffani, G. Minozzi, N. Morandi and A. Stella.
\newblock Application of genomics to problems in livestock production:
project ``ProZoo''.
\newblock {\em Plant \& Animal Genomes XIX}, San Diego (CA), USA, 2011.

\bibitem{barton2000}
N.H. Barton.
\newblock Gene hitch-hiking.
\newblock {\em Phil. Trans. R. Soc. London B}, 335:1553-1562, 2000.

\bibitem{biscarini2012}
F. Biscarini, E.L. Nicolazzi, A. Stella and the ProZoo Team.
\newblock Automated milk.recording systems: an experience in Italian
dairy cattle farms.
\newblock {\em $63^{rd}$ EAAP Book of Abstract}, 95, 2012.

\bibitem{cole2009}
J.B. Cole, P.M. VanRaden, J.R. O'Connell, C.P. Van Tassell, T.S. Sonstegard, R.D. Schnabel, J.F. Taylor and G.R. Wiggans.
\newblock Distribution and location of genetic effects for dairy traits.
\newblock {\em J. Dairy Sci.}, 92:2931--2946, 2009.

\bibitem{cole2011}
J. Cole, G. Wiggans, L. Ma, T. Sonstegard, T. Lawlor, B. Crooker, C. Van Tassell, J. Yang, S. Wang, L. Matukumalli and Y. Da.
\newblock Genome-wide association analysis of thirty one production,
health, reproduction and body conformation traits in contemporary US
Holstein cows.
\newblock {\em BMC Genomics}, 12(1):408, 2011.

\bibitem{cole2013}
J. Cole, K.P. Gaddis, J. Clay and C. Maltecca.
\newblock Genomic evaluation of health traits in dairy cattle.
\newblock {\em ICAR Technical Workshop 2013 and Health Conference},
29-31 May, Aarhus (Denmark), 2013.

\bibitem{agostino2002}
R.B. D'Agostino, J.M. Massaro and L.M. Sullivan.
\newblock Non-inferiority trials: design concepts and issues - the
encounters of academic consultants in statistics.
\newblock {\em Statis. Med.}, 22:169--186, 2002.

\bibitem{hansen2002}
M. Hansen, M.S Lund, M.K. S{\o}rensen and L.G. Christensen.
\newblock Genetic parameters of dairy character, protein yield, clinical
mastitis and other diseases in the Danish Holstein cattle.
\newblock {\em J. Dairy Sci.}, 85:445--452, 2002.

\bibitem{kolbehdari2008}
D. Kolbehdari, Z. Wang, J.R. Grant, B. Murdoch, A. Prasad, Z. Xiu, E.
Marques, P. Stothard and S.S. Moore.
\newblock A whole-genome scan to map quantitative trait loci for
conformation and functional traits in Canadian Holstein bulls.
\newblock {\em J. Dairy Sci.}, 91:2844--2856, 2008.

\bibitem{purcell2007}
S. Purcell, B. Neale, K. Todd-Brown, L. Thomas, M.A.R. Ferreira, D.
Bender, J. Maller, P. Sklar, P.I.W. de Bakker, M.J. Daly and P.C. Sham.
\newblock PLINK: a toolset for whole-genome association and
population-based linkage analysis.
\newblock {\em Am. J. Hum. Genet.}, 81:559--575, 2007.

\bibitem{R2013}
R Core Team.
\newblock A language and environment for statistical computing.
\newblock {\em R Foundation for Statistical Computing}, Vienna
(Austria), URL: ttp://www.R-project.org/, 2013.

\bibitem{luo1998}
Z.W. Luo.
\newblock Linkage disequilibrium in a two-locus model.
\newblock {\em Heredity}, 80:198--208, 1998.

\bibitem{minozzi2013}
G. Minozzi, E.L. Nicolazzi, A. Stella, S. Biffani, R. Negrini, B. Lazzari, P. Ajmone-Marsan and J.L. Williams
\newblock Genome-wide analysis of fertility and production traits in
Italian Holstein cattle.
\newblock {\em PLoS One}, 8(11):e80219, 2013.

\bibitem{pimentel2010}
E.C.G. Pimentel, S. Bauersachs, M. Tietze, H. Simianer, J. Tetens, G. Thaller, F. Reinhardt, E. Wolf and S. König.
\newblock Exploration of relationships between production and fertility
traits in dairy cattle via association studies of SNPs within candidate
genes derived by expression profiling.
\newblock {\em Anim. Genet.}, 42:251--262, 2010.

\bibitem{simianer1991}
H. Simianer, H. Solbu amd L.R. Schaeffer.
\newblock Estimated genetic correlations between diseases and yield
traits in dairy cattle.
\newblock {\em J. Dairy Sci.}, 74:4358--4365, 1991.

\bibitem{sullivan1994}
P.S. Sullivan, S.T. Grubbs, T.W. Olchowy, F.M. Andrews, J.G. White,
J.L. Catalfamo, P.A. Dodd and T.P. McDonald.
\newblock Bleeding diathesis associated with variant von Willebrand
factor in a Simmental calf.
\newblock {\em J. Am. Vet. Med. Assoc.}, 205(12):1763--1766, 1994.


\end{thebibliography}

\end{document}

